\documentclass[10pt]{article}

\usepackage{amsmath}
\usepackage{amssymb}

\usepackage{graphicx}

\usepackage{cite}

\usepackage{color} 

\usepackage[labelfont=bf,labelsep=period,justification=raggedright]{caption}

\makeatletter
\renewcommand{\@biblabel}[1]{\quad#1.}
\makeatother

\date{}

\begin{document}

\begin{flushleft}
{\Large
\textbf{Know the single-receptor sensing limit? Think again.}
}
\\ \ \\
Gerardo Aquino$^{1}$, Ned S. Wingreen$^{2}$, and Robert G. Endres$^{1,\ast}$
\\ \ \\
\bf{1} Department of Life Sciences \& Centre for Integrative Systems Biology and 
Bioinformatics, London, United Kingdom
\bf{2} Department of Molecular Biology, Princeton University, Princeton, New Jersey 08544, USA
\\ \ \\
$\ast$ E-mail: r.endres@imperial.ac.uk
\end{flushleft}

\section*{Abstract}
How cells reliably infer information about their environment is a fundamentally important question. While sensing and
signaling generally start with cell-surface receptors, the degree of accuracy with which a cell can measure external
ligand concentration with even the simplest device - a single receptor - is surprisingly hard to pin down. Recent studies
provide conflicting results for the fundamental physical limits. Comparison is made difficult as different studies either suggest
different readout mechanisms of the ligand-receptor occupancy, or differ on how ligand diffusion is implemented. 
Here we critically analyse these studies and present a unifying perspective on the limits of sensing, with wide-ranging 
biological implications.

\section*{Introduction}
In 1977, physicists Howard Berg and Edward Purcell published their results on the 
fundamental biological problem of sensing \cite{berg77}. The question they 
addressed was how accurately a biological cell, viewed as a tiny measurement device, 
can sense its chemical environment using cell-surface receptors. \textcolor{black}{The 
paper is not only highly cited, but, more importantly, a large fraction of the 
citations stems from the last ten years, demonstrating how far ahead of its time the study was.} 
In essence, the message of the paper was simple: sensing in the
microscopic world boils down to counting molecules, which arrive at
the cell surface by diffusion. Humans encounter a
similar limit when we try to see in the near dark as our photoreceptors count 
single photons \cite{rieke98}. Berg and Purcell's paper has influenced many 
fields of quantitative biology, including nutrient scavenging 
\cite{dusenberg98,bialek05,endres08}, mating \cite{noa12}, signal transduction 
\cite{bialek05,hu11}, gene regulation \cite{tkacik09}, cell division 
\cite{howard12,halatek12,kerr06}, and embryonic development \cite{gregor07}. 
While there is no disagreement on the importance of knowing the fundamental 
physical limits of sensing, there has been disagreement on what this limit is, even 
for a single receptor. The analysis here interprets and unifies these studies
to yield a coherent picture of the limits of sensing.

\section*{Overview}
\begin{figure}[t]
\includegraphics[width=14cm]{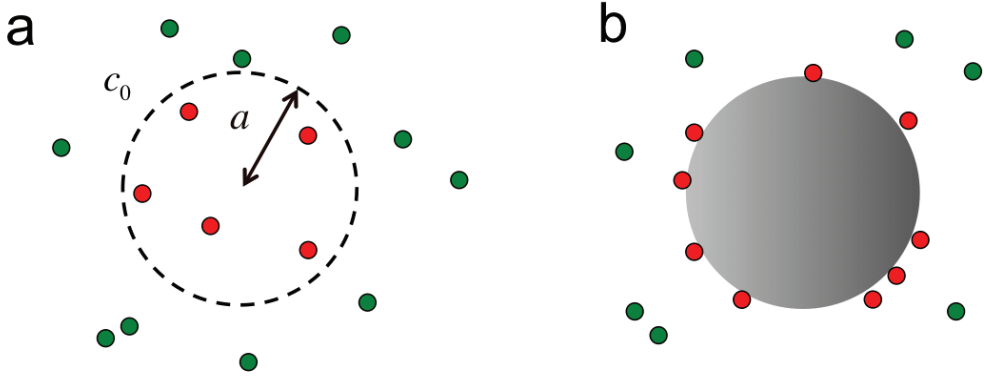}
\caption{
{\bf Simple measurement devices for concentration.}  (a) The Perfect Monitor
is permeable to ligand molecules and estimates the concentration $c_0$ by
counting the molecules in its volume during time $T$. (b) The Perfect Absorber estimates 
the ligand concentration from the number of molecules incident on its surface
during time $T$.
}
\label{Fig1}
\end{figure}

To introduce the topic and to build intuition, we follow Berg
and Purcell \cite{berg77} and begin with simple models for measuring ligand
concentration $c_0$. The first is the Perfect Monitor \cite{berg77}. 
This model assumes a permeable sphere of radius $a$,
capable of counting the number of molecules $N$ inside its volume (Fig. \ref{Fig1}a).
For concreteness, the sphere might represent a bacterial cell. Since the molecules 
diffuse independently, finding a molecule in one small 
volume element is independent of finding another one in a different small volume element,
and so the number of molecules $N$ will be Poisson distributed. Since for the Poisson distribution  
the variance equals the mean, i.e. $\delta N^2=\bar N$ (omitting ensemble-averaging 
brackets for simplicity of notation), we obtain for a single measurement (``snapshot'')
\begin{equation}
\frac{\delta c^2}{c_0^2}=\frac{\delta N^2}{\bar N^2}=\frac{1}{\bar N}=
\frac{1}{c_0V},\label{Eq1}
\end{equation}
where $c_0$ is a fixed, given ligand concentration and $V$ is the volume of the monitoring sphere.
However, if we assume the Perfect Monitor has some time $T$ available to make
a measurement, the uncertainty in the estimate of the true ligand concentration can be 
further reduced. In time $T$, the Perfect Monitor can make approximately $M\sim T/\tau_D$ statistically
independent measurements, where $\tau_D\sim a^2/D$ is the diffusive turnover time for the 
molecules inside the sphere. This leads to the reduced uncertainty
\begin{equation}
\frac{\delta c^2}{c_0^2}=\frac{1}{M\bar N}=
\frac{1}{(T/\tau_D)c_0V}\sim\frac{1}{Dac_0T},\label{Eq2}
\end{equation}
where we neglect prefactors for this heuristic derivation. (The exact result is $3/(5\pi Dac_0T)$, 
which can be derived by considering autocorrelations of the molecules inside the volume \cite{berg77}.)

However, the Perfect Monitor is not the best one can do. A more accurate estimate
can be made if each ligand molecule is only measured {\it once} rather than being
allowed to diffuse in and out of the sphere. Thus, we consider 
a perfectly absorbing sphere \cite{endres08}, 
estimating concentration from the number of absorbed ligand molecules 
$N_T$ in time $T$, and find (Fig. \ref{Fig1}b)
\begin{equation}
\frac{\delta c^2}{c_0^2}=\frac{1}{N_T}=\frac{1}{4\pi Dac_0T}<\frac{3}{5\pi Dac_0T}\label{Eq3}
\end{equation}
This Perfect Absorber is thus more accurate than the Perfect
Monitor (and even more so for spatial gradient sensing by almost a factor of 10) 
\cite{endres08}. \textcolor{black}{This result contrasts with Berg and Purcell's 
original suggestion that rebinding previously measured ligand molecules 
does not increase the uncertainty in measurement \cite{berg77}.
However, one of their many key insights was that a sphere with many absorbing 
patches for ligand is nearly as good at sensing as a fully absorbing sphere, making room 
for multiple receptor types with different ligand specificity without sacrificing
much accuracy.}

\begin{figure}[t]
\includegraphics[width=14cm]{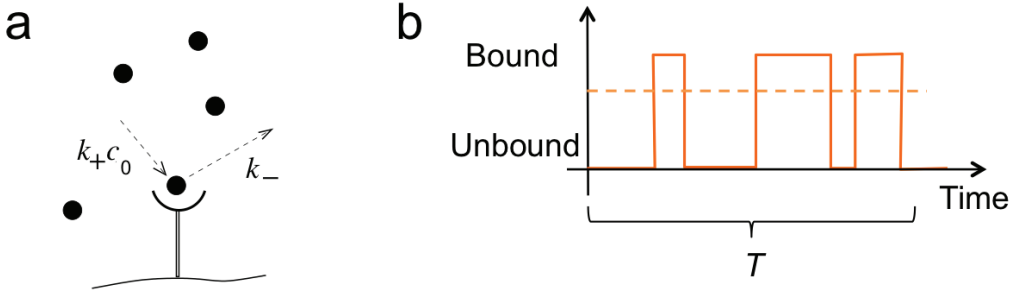}
\caption{
{\bf Measuring ligand concentration with a single receptor.} (a) A
receptor binds ligand with rate $k_+c_0$ when unbound, and unbinds
ligand when bound with rate $k_-$. (b) Time series of receptor
occupancy during time interval $T$. Berg and Purcell considered the accuracy
obtained by taking the average (dashed horizontal line).
}
\label{Fig2}
\end{figure}

\section*{Single receptor without ligand rebinding}
The single receptor is the simplest measurement device and thus needs to be
thoroughly understood. Unfortunately, different approaches to estimating its sensing
accuracy have resulted in significant discrepancies. We first disregard the effects of 
diffusion and rebinding of previously bound ligands, and just consider ligand binding and unbinding.

Consider the receptor shown in Fig. \ref{Fig2}a, which binds 
ligand with rate $k_+c_0$ when unbound and unbinds ligand with rate $k_-$ when bound. The
probability of being bound is then $p=c_0/(c_0 +K_D)$ with $K_D=k_-/k_+$ 
the ligand dissociation constant. A potential time series of receptor occupancy 
$\Gamma(t)$ during time $T$ is illustrated in Fig. \ref{Fig2}b. Berg and Purcell argued
that the best a cell can do to estimate the ligand concentration is to average the 
occupancy $\Gamma(t)$ over time. For such an average, the variance $\delta \Gamma^2$ was 
derived from the autocorrelations of occupancy, leading to
the relative uncertainty in estimating the ligand concentration 
\begin{equation}
\frac{\delta c^2}{c_0^2}=\left(c_0\frac{\partial p}
{\partial c}\right)^{-2}\delta \Gamma^2\label{Eq4},
\end{equation}
where the derivative $\partial p/\partial c$ is the gain or amplification. Naively, one could
be tempted to set $\delta \Gamma^2=p(1-p)$ equal to the variance of a
Bernoulli random variable (binomial trials). However, this would correspond to the
uncertainty in the concentration estimate following a single instantaneous observation 
of the state of the receptor, or equivalently to the frequency integral of the noise 
power spectrum $\delta \Gamma^2=\int d\omega /(2\pi) S_\Gamma(\omega)$
(see Supplementary Information for details). \textcolor{black}{This snapshot limit can be improved 
if we assume $T{>\!\!>}1/(k_+c_0)+1/k_-$, i.e.
that the receptor is allowed to average over a time $T$ much larger than the correlation 
time of ligand binding and unbinding.} In this case, one can take the low-frequency limit
$\delta \Gamma^2\approx S_\Gamma(\omega=0)/T$ instead, and Eq. \ref{Eq4} leads to
\textcolor{black}{the Berg-Purcell limit for a single receptor}
\begin{equation}
\frac{\delta c^2}{c_0^2}=\frac{2\tau_b}{T p}=\frac{2}{\bar N}
\rightarrow\frac{1}{2Dac_0(1-p)T}\label{Eq5},
\end{equation}
where $\tau_b=1/k_-$ is the average duration of a bound interval. The simple
formulation as $2/\bar N$ follows because the average number of binding and unbinding 
events in time $T$ is $\bar N=T/(\tau_b+\tau_u)$, where $\tau_u=(k_+c_0)^{-1}$ is the
average duration of an unbound interval. The final result in Eq. \ref{Eq5} follows from 
detailed balance for diffusion-limited binding.

But is it true that averaging receptor occupancy is the best way to estimate concentration?
More recently a limit lower than Eq. \ref{Eq5} was found by applying
maximum-likelihood estimation to a time series $\Gamma(t)$ of receptor
occupancy \cite{endres09b}. Here the probability $P(\Gamma,c)$ of observing a time series 
$\Gamma$ is maximised with respect to the concentration $c$. The resulting 
best estimate of the concentration comes only from the unbound intervals, since 
only they depend on the rate of binding and thus on the ligand concentration. To obtain
a lower limit on the uncertainty the Cram\'er-Rao bound \cite{cover06}
can be used, leading to
\begin{equation}
\frac{\delta c^2}{c_0^2}\geq\frac{1}{c_0^2I(c_0)}\rightarrow\frac{1}{N}\label{Eq6},
\end{equation}
where $I(c_0)=-\partial^2\ln(P)/\partial c^2$ is the Fisher information evaluated at $c_0$  
and averaged over all trajectories with the same $N$ (when employing  maximum-likelihood
estimation it is easier to work with a fixed number of binding/unbinding events $N$
than a fixed time $T$). The limit on the right-hand side of Eq. \ref{Eq6}
is obtained for long time series for which the inequality becomes an equality. 
\textcolor{black}{Note, however,
that a slightly sharper bound $1/(N-2)$ can be obtained when using a further improved
estimator (see Supplementary Information).}
Eq. \ref{Eq6} shows that the uncertainty in Eq. \ref{Eq5} can be reduced 
by a factor of two. This is because only unbound intervals carry information about 
the ligand concentration. In contrast, the bound intervals only increase the 
uncertainty and hence are discarded by the maximum-likelihood procedure. 

What does the maximum-likelihood result imply about tuning receptor parameters to minimise the
uncertainty? The minimal uncertainty is obtained for $N_\text{max}$, the maximal 
number of binding events provided by very fast unbinding ($k_-\rightarrow \infty$). 
This ideal limit corresponds to the Perfect Absorber from Eq. \ref{Eq3} as every binding 
event is counted. (However, the increased accuracy comes at the expense of 
specificity as any ligand molecule dissociates immediately and 
hence different ligand types cannot be differentiated.) Maximum-likelihood estimation  
can also be extended to ramp sensing (temporal gradients) \cite{mora10} 
and multiple receptors \cite{mortimer10,hu10}.

\begin{figure}[t]
\includegraphics[width=10cm]{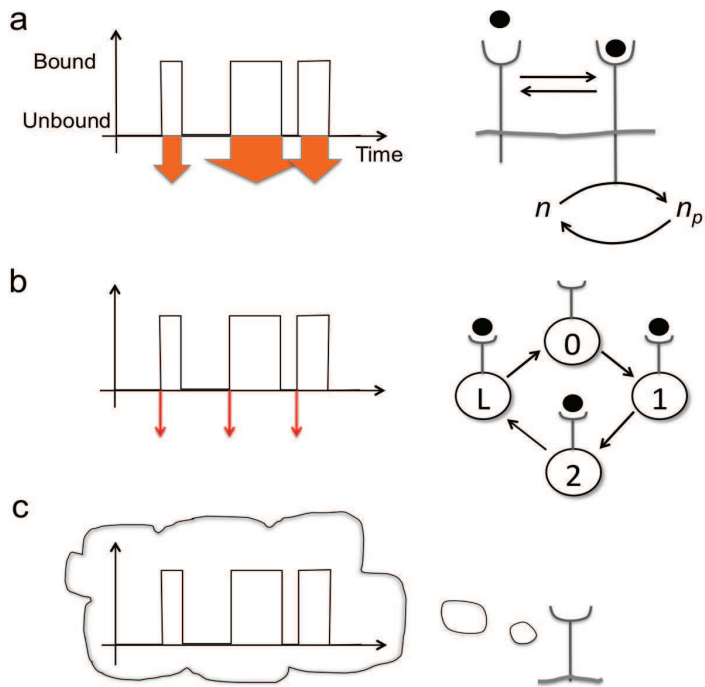}
\caption{
{\bf Three schemes of receptor readout.} (a) Integrating receptor, 
which signals while ligand bound (left) \cite{berg77}. For example, 
the active receptor might phosphorylate a protein with concentration 
$n$. The concentration of the phosphorylated protein is $n_p$ (right). 
(b) Alternatively, the receptor may signal in generic bursts 
at onset of ligand binding (left) \cite{endres09b}. This scheme can be 
implemented by an energy-driven cycle of $L$ active/bound receptor 
conformations, which reduces variability (right) \cite{lang14}.
(c) A receptor could also retain a memory of previous binding and 
unbinding events, potentially improving its accuracy of sensing 
\cite{aquino14}.
}
\label{Fig3}
\end{figure}

Adding a downstream signaling molecule cannot increase the accuracy of sensing, 
in fact this only adds noise. 
For example, consider an integrating receptor \`a la Berg and Purcell, 
which signals while being ligand bound  (Fig. \ref{Fig3}a) \cite{mehta12}. 
In this simple network a downstream signaling molecule with concentration $n$ is 
phosphorylated by ligand-bound receptors with the phosphorylated concentration given by 
$n_p$ with lifetime $\tau$ (beyond this time the protein converts back to the 
unphosphorylated form). \textcolor{black}{Now, instead of taking the snapshot limit, i.e. the total 
variance $\delta n_p^2$, we time average to reduce the uncertainty. Specifically, let us assume a 
long averaging time, that is $T>\!\!>\tau>\!\!>1/(k_+c_0)+1/k_-$, allowing us to use again the 
low-frequency limit of the corresponding power spectrum.} 
We then obtain (see Supplementary Information for details)
\begin{eqnarray}
\frac{\delta c^2}{c_0^2}&=&\left[{\frac{2}{\bar N_\tau}}+
{\frac{2}{\bar n(1-p)^2}}\right]
{\frac{\tau}{T}}\label{Eq7a}\\
&=&\underbrace{\frac{2}{\bar N}}_\text{BP limit}+
\underbrace{\left[\frac{2}{\bar n(1-p)^2}\right]}_\text{Poisson-like}
\underbrace{\frac{\tau}{T}}_\text{time ave}.\label{Eq7b}
\end{eqnarray}
Eq. \ref{Eq7b} shows that by integrating receptor output one cannot do better than the 
Berg-Purcell limit, given by the first term. 
The second term represents additional Poisson-like noise from number fluctuations of the 
signaling molecule due to imperfect averaging \cite{paulsson05}. For $T{>\!\!>}\tau$ the
Berg-Purcell limit is approached from time averaging this noise. While we focus here 
on averaging in time of stationary stimuli, non-stationary ligand concentrations 
may be more accurately sensed via non-uniform time averaging, requiring appropriately
designed signaling cascades \cite{govern12}.

\textcolor{black}{There has been some confusion about whether sensing actually costs energy.
On the one hand, C. H. Bennett pointed out long ago that sensing does not need to cost
if done reversibly (and hence extremely slowly) \cite{bennett73}. On the other hand, cells obviously 
consume energy, e.g. using ATP to phosphorylate proteins. In other words, what energy cost is 
actually necessary for performing a measurement? As stressed by the authors in \cite{mehta12}
the process of sensing in terms of ligand-receptor binding does not need to cost energy if done 
using an equilibrium receptor in the spirit of Berg and Purcell. However, to accurately infer the external
ligand concentration, the cell needs to time average, which cannot be done without consuming
energy. This is in line with the Landauer erasure principle \cite{landauer61}, which 
predicts a lower theoretical limit of energy consumption of a computation. In essence, to keep a 
record of the past for averaging, old information needs to be erased and time-reversal symmetry 
broken \cite{govern14}. Time averaging can be implemented by phosphorylation of a 
downstream protein: when the receptor is bound it phosphorylates and when unbound it 
dephosphorylates. Since these are energetically driven reactions the reverse reaction, e.g. 
dephosphorylation by a bound receptor is extremely unlikely, and time averaging is very 
efficient. The issue of the cost was avoided in Berg and Purcell's analysis by providing an effective averaging 
time $T$ without specifying how this averaging is achieved.}

How is the maximum-likelihood result, Eq. \ref{Eq6}, useful? The maximum-likelihood result 
makes interesting predictions about sophisticated sensing strategies cells
might employ. For example, to implement maximum likelihood \textcolor{black}
{in the fast unbinding limit} a receptor should only signal upon a ligand-binding event 
as illustrate in Fig. \ref{Fig3}b (thin arrows), rather than continuously signaling 
while ligand is bound \textcolor{black}{(see \cite{micali15} for further discussion)}. How can the cell 
achieve such short and well-defined signaling durations? Reducing variability and 
achieving determinism requires energy consumption and irreversible cycles \cite{lang14}. 
Examples may include ligand-gated ion channels \cite{csanady10} and single-photon 
responses in rhodopsin of rod cells \cite{doan06}. 

Maximum likelihood provides another valuable insight - it shows that information
from an estimate and memory from a prior are equivalent, and both can
contribute to lowering the uncertainty (Fig. \ref{Fig3}c). This kind of
receptor ``learning'' from past estimates can be implemented using the
Bayesian Cram\'er-Rao bound for the uncertainty. Using prior 
information $I(\lambda)$, one obtains \cite{aquino14}
\begin{equation}
\frac{\delta c^2}{c_0^2}=-\frac{1/c_0^2}
{\underbrace{I(c_0)}_\text{Fisher info.}+
\underbrace{I(\lambda)}_\text{prior}}=\frac{1}{2N}\label{Eq8},
\end{equation}
assuming the prior had variance $1/N$, identical to the actual measurement. 
Importantly, memory can even help in fluctuating environments if a filtering scheme 
is implemented by the cell \cite{aquino14}: if the environment fluctuates weakly 
and/or with long temporal correlations, memory improves precision significantly. 
If, on the other hand, the environment fluctuates very strongly and/or without any 
correlations, the cell can still rely on the current measurement (and disregard memory). 
A form of memory is implemented by receptor methylation in 
bacterial chemotaxis \cite{koshland82}, and in principle memory could be implemented by
any slow process in the cell, e.g. the expression of LacY permease in enzyme 
induction in the lac system \cite{ozbudak04}, or the remodelling of the actin cortex 
in eukaryotic chemotaxis \cite{cooper12,westendorf13}.

\section*{Single receptor with ligand rebinding}

So far we have neglected the possibility of rebinding by previously bound ligands. 
In fact, the role of ligand rebinding in the
accuracy of sensing is a tricky issue, because rebinding can
introduce non-trivial correlations between binding events. In practice, these
correlations can only be included approximately in analytical calculations, and so the
question is how to proceed. Originally Berg and Purcell made the
reasonable suggestion that a molecule that fails to bind to a receptor may
return to the receptor by diffusion and rebind, and that this effect 
may be included by considering diffusion-limited binding with a
renormalised receptor size \cite{berg77}. However, the question is how to
formally separate ligand binding and unbinding from ligand
diffusion.

Bialek and Setayeshgar addressed this problem by coupling ligand-receptor binding 
and unbinding to the diffusion equation \cite{bialek05}. 
Assuming that the averaging time is long compared to the typical binding and 
unbinding time, the low-frequency limit can be used. This results in
\begin{equation}
\frac{\delta c^2}{c_0^2}={\frac{2}{k_+c_0(1-p)T}}
+{\frac{1}{\pi Dac_0T}}\label{Eq9}
\end{equation}
with $a$ now the size of the receptor. Equation \ref{Eq9} indicates noise
contributions from two independent sources. According to Ref. \cite{bialek05}, 
the first term represents binding and unbinding noise and depends on the rate
parameters, while the second depends on diffusion and was interpreted as
a Berg-Purcell-like noise floor. However, we argue for a different interpretation:
For diffusion-limited binding, the first term in Eq. \ref{Eq9} is not zero, but 
rather $k_+$ needs to be set to a Kramer-like expression, which is proportional
to the diffusion constant \cite{berezh14} and an Arrhenius factor at most equal to one 
\cite{pollak05}. Due to their dependence on the diffusion constant, both terms can be 
combined \cite{endres09}. Indeed, for 
diffusion-limited binding it is the first, not the second term of Eq. \ref{Eq9} that captures the 
Berg-Purcell limit. Since Berg and Purcell did not consider rebinding by diffusion, the 
second term constitutes increased noise due to a rebinding correction that does not arise in Berg and Purcell's 
derivation \cite{berg77}. Bialek and Setayeshgar also applied their method to multiple 
receptors, and showed that the second term can introduce correlations among receptors, 
as ligand unbinding at one receptor can lead to re-binding at another nearby receptor 
\cite{bialek08,endres09}. \textcolor{black}{Hence, while multiple independent receptors
allow for spatial averaging \cite{berez13}, mutual rebinding among different receptors 
by diffusion increases the uncertainty of sensing.}

More recently, Kaizu {\it et al.} readdressed this problem \cite{kaizu14} by applying a 
formalism developed by Agmon and Szabo for diffusion-influenced reactions \cite{agmon90}. 
By calculating survival probabilities of bimolecular reactions with a number of 
simplifying assumptions (see below), they obtained for the relative uncertainty
of a single receptor
\begin{equation}
\frac{\delta c^2}{c_0^2}={\frac{2}{k_+c_0(1-p)T}}
+{\frac{1}{2\pi Dac_0(1-p)T}}\label{Eq10}.
\end{equation}
Similar to Bialek and Setayesghar, there are two noise contributions with 
the first terms in Eq. \ref{Eq9} and \ref{Eq10} formally identical. The second term is, however, different. 
While the lost factor 2 in the second term in Eq. \ref{Eq10} can be traced to different definitions of the 
receptor geometry \textcolor{black}{(cubic in Eq. \ref{Eq9} and spherical in Eq. \ref{Eq10})}, the 
factor $1-p$ in Eq. \ref{Eq10} is missing from Eq. \ref{Eq9}. Due to this factor, both terms of the 
uncertainty in Eq. \ref{Eq10} diverge 
if the receptor is fully bound on average ($p=1$), while in Eq. \ref{Eq9} only the 
first term diverges. Unlike Eq. \ref{Eq9} Kaizu {\it et al.} made the additional assumption that 
during a bound interval the external ligand equilibrates. As a result, an unbound ligand molecule 
cannot diffuse away and rebind at a later time with another ligand bound in between. However, using 
exact simulations they showed that such delayed rebinding is only a minor effect under biologically relevant 
conditions. Hence, as the factor $1-p$ in the second term of Eq. \ref{Eq10} also appears 
in the Berg-and-Purcell limit, Eq. \ref{Eq5}, Kaizu {\it et al.} argue that their result is more
accurate than Eq. \ref{Eq9}.

\textcolor{black}{However, we propose a slightly different interpretation of Eq. \ref{Eq10}. Similar 
to \cite{bialek05} we suggest that the second term is not the Berg-Purcell limit (Eq. \ref{Eq5}) 
for diffusion-limited binding since the first term captures the Berg-Purcell limit \cite{kaizu14}.} 
As described above, for diffusion-limited binding the first term cannot be neglected. 
This aside, how may the factor $1-p$ in
the second term be interpreted? In Kaizu {\it et al.}'s derivation, diffusion means that a ligand molecule 
enters the ligand pocket of a receptor without actually binding (hence the factor $1-p$ in the second 
term since they assume only an unbound receptor can be approached by a ligand molecule). 
\textcolor{black}{In Bialek and Setayeshgar's derivation, no such factor appears as the second term describes fluctuations 
in ligand concentration simply in the vicinity of the receptor. Can further insight into the effects of diffusion be 
obtained by yet an alternative method?}

Maximum-likelihood estimation can also be applied to a receptor with ligand diffusion, \textcolor{black}{albeit 
only in a special case.} The probability of observing a time series 
of receptor occupancy of $N$ binding and unbinding events can be formally written down even with
diffusion \cite{endres09b}. However, the rate of binding will depend on the current ligand concentration, 
which is influenced by the history of all previous binding and unbinding events (even 
before the first recorded binding event). To estimate the uncertainty, the Cram\'er-Rao 
bound can be applied but cannot be evaluated exactly. Nevertheless, for fast diffusion 
or slow binding an approximate expression can be derived \textcolor{black}{for both 2D and 3D 
(see Supplementary Information)}

\begin{equation}
\frac{\delta c^2}{c_0^2}\approx\frac{1}{N}\left(1+2\frac{\Delta c}{c_0}\right)
=\frac{1}{k_+c_0(1-p_{1/2})T}+
\left\{\begin{array}{ll}
\frac{\ln(4\pi D/(k_+c_0a^2))}{2\pi Dc_0(1-p_{1/2})T} & \mbox{for 2D}\\
\frac{1}{\pi Dac_0(1-p_{1/2})T} & \mbox{for 3D}
\end{array}\right.\label{Eq11}.
\end{equation}
\textcolor{black}{In Eq. \ref{Eq11} the average local ``excess'' ligand concentration $\Delta c$ 
due to previous binding and unbinding events is $k_+c_0/(4\pi D)\cdot\ln[4\pi D/(k_+c_aa^2)]$ in 2D and 
$k_+c_0/(2\pi Da)$ in 3D (for the derivation half occupancy $p_{1/2}=1/2$ is required). 
The ratios in the excess concentration reflect the competition between rebinding and diffusion. 
As expected, in 2D this concentration decays more slowly to zero with increasing diffusion constant 
than in 3D, and also the spatial dependence on the receptor size is weaker in 2D than in 3D.} 

\textcolor{black}{Coming back to the different receptor models with diffusion, the first term of 
Eq. \ref{Eq11} produces exactly half the uncertainty of the first terms of Bialek and Setayeshgar 
(Eq. \ref{Eq9}) and Kaizu {\it et al.} (Eq. \ref{Eq10}) by utilizing only the unbound time intervals. 
However, due to factor $1-p$ the second term of Eq. \ref{Eq11} resembles the second term of 
Kaizu {\it et al.} (both Eq. \ref{Eq10} and Eq. \ref{Eq11} for 3D use a spherical receptor). This 
suggests that Kaizu {\it et al.} is the correct result for the accuracy of sensing by time averaging, 
while Eq. \ref{Eq11} is the more accurate result when using maximum-likelihood estimation.}

\begin{figure}[t]
\includegraphics[width=12cm]{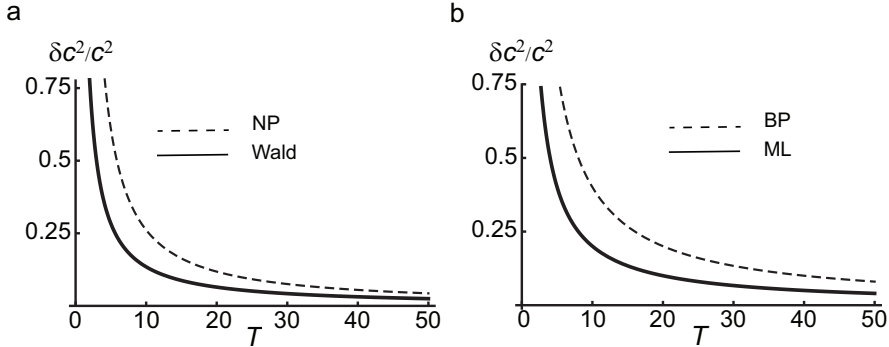}
\caption{{\bf Comparison of decision-making algorithms and fixed-time algorithms.}
\textcolor{black}{(a) Decision-making algorithms: Wald algorithm (solid curve) has lower uncertainty than 
fixed-time log-likelihood ratio estimation based on the Neyman-Pearson (NP) lemma (dashed 
curve). Uncertainty is calculated by converting decision error into variance.
(b) Uncertainty estimates based on direct measurement of ligand concentration in a fixed amount of 
time: Maximum-likelihood (ML) estimation (solid curve) based on the Cram\'er-Rao bound of 
Fisher information has only half the uncertainty of the Berg-Purcell (BP) limit (dashed curve) 
for the standard error of the mean concentration. For further details see Supplementary 
Information.}}
\label{Fig4}
\end{figure}

\section*{Single receptor as a decision maker}
All the above approaches considered the accuracy based on a fixed measurement time (or number
of binding and unbinding events). However, similar to humans, cells might follow a different strategy 
and approach a problem from a decision-making perspective \cite{perkins09,siggia13}: 
either deciding based on existing information or waiting to accumulate more data. 

Recently, Siggia and Vergassola considered decision making in the context of cells, 
proposing that the above maximum-likelihood estimate can further be improved in this way \cite{siggia13}. 
The simplest implementation of a decision-making strategy is the so-called Wald algorithm \cite{wald45}. 
For a single receptor, the Wald algorithm requires calculating the ratio $R$ of the likelihoods of 
the time series of binding and unbinding events of a receptor $\Gamma$ (data), conditioned to either 
of two hypothesised values of external ligand concentration 
\begin{equation}
R=\frac{P(\Gamma|c_1)}{P(\Gamma|c_2)}. \label{EqR}
\end{equation}
The cell then concludes that the ligand concentration is $c_1$ if $R\geq H_1$, that the ligand concentration 
is $c_2$ if $R\leq H_2$, or keeps collecting data if $H_1<R<H_2$.  $H_1$ and $H_2$ are thresholds  
that set the probability of error, i.e. concluding the concentration is $c_1$ if the true concentration is
$c_2$ and vice versa. 
This algorithm, by not having a fixed-time constraint, can be shown to be optimal, i.e. the average time 
to make a decision between the two options is shorter than provided by any other algorithm with the 
same accuracy (decision-error probability).

How can decision making be compared with maximum-likelihood estimation and the Berg-Purcell limit? 
Siggia and Vergassola suggested a fixed-time log-likelihood-ratio estimation \`a la Eq. \ref{EqR}
based on the Neyman-Pearson lemma. Due to the fixed-time constraint the Neyman-Pearson
algorithm is in spirit similar to maximum-likelihood estimation. Siggia and Vergassola showed 
that the Wald algorithm leads to a shorter decision-making time, on average, than the Neyman-Pearson
algorithm, and so suggested that the Wald algorithm provides the ultimate limit for sensing.
 
The result for the Wald algorithm indeed shares properties with the maximum-likelihood estimate and the 
Berg-Purcell limit. All three reveal a dependence of the measurement (decision) 
time on the inverse of the square of the difference of concentration (i.e. $\Delta c^2$). Although 
no decision making is involved in maximum-likelihood estimation or the Berg-Purcell limit, 
one can still conclude that concentrations $c_1$ and $c_2$ can be distinguished if the measurement 
uncertainty is smaller than the difference $\delta c^2 < (c_2 -c_1)^2$, and that, assuming either 
$c_1$ or $c_2$ as the true value, an incorrect decision occurs if measurement returns a value closer 
to the wrong concentration. This way a decision error can be converted into a type of measurement 
uncertainty and vice versa (see Supplementary Information for details).

\textcolor{black}{Fig. 4a shows the thus derived uncertainty in measuring a ligand concentration by
the Wald algorithm and the Neyman-Pearson lemma as a function of average measurement time. For comparison the 
maximum-likelihood estimate and the Berg-Purcell limit are shown in Fig. 4b.} 
However, since the fixed-time likelihood algorithms (Neyman-Pearson lemma 
and maximum-likelihood estimate) do not agree, it is difficult to directly compare Wald with the Berg-Purcell 
limit. After all, Wald and Neyman-Pearson algorithms are about hypothesis testing and discrimination, 
while time averaging (Berg-Purcell) and maximum likelihood are about estimation.

What types of algorithm are cells actually implementing? Consider chemotaxis in the bacterium 
{\it Escherichia coli} as a prototypical example of chemical sensing. Downstream signaling, 
especially slow motor switching, could provide a time scale for Berg-Purcell-type time averaging. 
In contrast, biological systems with hysteresis, that is two different thresholds for activation 
and deactivation of the downstream pathway, may implement a type of decision-making algorithm. 
The classical example is the lactose utilisation system in {\it E. coli}, which can be stimulated 
by the non-metabolisable 'gratuitous' inducer TMG \cite{ozbudak04,perkins09}. When TMG is high enough 
enzymes of the lac system become induced. Once induced, however, the TMG level must be reduced below 
a much lower threshold in order to uninduce the lac system.

\section*{Outlook}
While the question of the physical limits of sensing has been around for decades, only over the
last few years has the importance of this question become clear and its predictions testable by
quantitative experiments \cite{little13,tweedy13}. While current work is mostly about chemical sensing, 
the limits of sensing other stimuli, such as substrate stiffness during durotaxis 
\textcolor{black}{(or temperature, pH, particles, and combinations of them, etc.) may be next. 
For such measurements, the role of domain size and spatial dimension are interesting 
questions. Measurements are often done inside a cell, on 2D surfaces, or along 1D DNA
molecules, and 
correlations due to rebinding depend on these parameters \cite{bicknell15}.}

The question of the limits of sensing has also opened up completely new directions, including the 
role of active, energy-consuming sensing strategies \cite{murugan12,mehta12,skoge13,lang14}, and 
hence the importance of nonequilibrium-physical processes in cell biology. This then connects to 
the Landauer limit of information erasure and cellular computation in general \cite{toyabe10,berut12}. 
\textcolor{black}{In this area, important questions are linking information theory, statistical inference, 
and thermodynamics e.g. in order to produce generalized second laws \cite{barato14}. Additionally, analysis
may move away from only considering receptors to considering receptors and their downstream signaling pathways, 
and questions of optimal resource allocation in such pathways emerge \cite{govern14}.}

Other areas of study have started to benefit from this work as well, such as gene regulation. 
For instance, why do cells often use bursty frequency modulation of gene expression under stress 
and in development \cite{levine13}? This may either reflect a need for accurately sensing and 
monitoring chemical cues, or simply enhance robustness, e.g. similar to when information is 
transmitted between neurons by action potentials. The questions whether cells sense at the physical 
limit and if so, how they reach it, and how to design experiments to answer these questions will 
occupy us for a while.

\section*{Acknowledgments}
GA and RGE thankfully acknowledge financial support by the Leverhulme-Trust Grant N. RPG-181. 
RGE was also supported by the European Research Council Starting-Grant N. 280492-PPHPI. 
NSW was supported by National Science Foundation Grant PHY-1305525. \textcolor{black}{We also
would like to thank an anonymous referee for his valuable comments on the Cram\'er-Rao bound.} 


\bibliography{SIngleReceptor_FINAL2.bib}
\title{Supplementary Information to:\\ Know the single-receptor sensing limit? Think again.}

\author{Gerardo Aquino$^1$, Ned S. Wingreen$^2$ and Robert G.  Endres$^1$}

\newcommand{\dT}{\Delta t}
\newcommand{\bigtiangledown}{\grad}
\newcommand{\Sss}{\scriptscriptstyle}
\newcommand{\Ss}{\scriptstyle}
\newcommand{\D}{\dysplaystyle}
\newcommand{\T}{\textstyle}
\newcommand{\e}{{\rm e}}
\newcommand{\veps}{\varepsilon}
\newcommand{\epss}{\varepsilon_{\sigma}}
\newcommand{\epsv}{\varepsilon_{V}}
\newcommand{\lgl}{\langle}
 \newcommand{\rgl}{\rangle}
 \newcommand{\gammaa}{\gamma_{\alpha}}

\newcommand{\Vh}[1]{\hat{#1}}
\newcommand{\Aa}{A^1_{\epsilon}}
\newcommand{\Ab}{A^{\epsilon}_L}
\newcommand{\Ae}{A_{\epsilon}}

\newcommand{\finn}[1]{\phi^{\pm}_{#1}}
\newcommand{\ea}{e^{-|\alpha|^2}}
\newcommand{\eb}{\frac{e^{-|\alpha|^2} |\alpha|^{2 n}}{n!}}
\newcommand{\ebbb}{\frac{e^{-3|\alpha|^2} |\alpha|^{2 (l+n+m)}}{l!m!n!}}
\newcommand{\ass}{\alpha}
\newcommand{\as}{\alpha^*}
\newcommand{\fb}{\bar{f}}
\newcommand{\gb}{\bar{g}}
\newcommand{\la}{\lambda}
 \newcommand{\sz}{\hat{s}_{z}}
\newcommand{\sy}{\hat{s}_y}
\newcommand{\sx}{\hat{s}_x}
\newcommand{\sio}{\hat{\sigma}_0}
\newcommand{\six}{\hat{\sigma}_x}
\newcommand{\siz}{\hat{\sigma}_{z}}
\newcommand{\siy}{\hat{\sigma}_y}
\newcommand{\vhsig}{\vec{\hat{\sigma}}}
\newcommand{\hsig}{\hat{\sigma}}
\newcommand{\hH}{\hat{H}}
\newcommand{\hU}{\hat{U}}
\newcommand{\hA}{\hat{A}}
\newcommand{\hB}{\hat{B}}
\newcommand{\hC}{\hat{C}}
\newcommand{\hD}{\hat{D}}
\newcommand{\hV}{\hat{V}}
\newcommand{\hW}{\hat{W}}
\newcommand{\hK}{\hat{K}}
\newcommand{\hX}{\hat{X}}
\newcommand{\hM}{\hat{M}}
\newcommand{\hN}{\hat{N}}
\newcommand{\te}{\theta}
\newcommand{\vze}{\vec{\zeta}}
\newcommand{\vet}{\vec{\eta}}
\newcommand{\vx}{\vec{\xi}}
\newcommand{\vc}{\vec{\chi}}
\newcommand{\hro}{\hat{\rho}}
\newcommand{\vro}{\vec{\rho}}
\newcommand{\hR}{\hat{R}}
\newcommand{\half}{\frac{1}{2}}
\renewcommand{\d}{{\rm d}}
\renewcommand{\top }{ t^{\prime } }
\newcommand{\oz}{{(0)}}
\newcommand{\sint}{{\rm si}}
\newcommand{\cint}{{\rm ci}}
\newcommand{\de}{\delta}
\newcommand{\ep}{\varepsilon}
\newcommand{\De}{\Delta}
\newcommand{\eps}{\varepsilon}
\newcommand{\si}{\hat{\sigma}}
\newcommand{\om}{\omega}
\newcommand{\tr}{{\rm tr}}
\newcommand{\ha}{\hat{a}}
\newcommand{\gam}{\gamma ^{(0)}}
\newcommand{\pe}{\prime}
\newcommand{\BEQ}{\begin{equation}}
\newcommand{\EEQ}{\end{equation}}
\newcommand{\BEA}{\begin{eqnarray}}
\newcommand{\EEA}{\end{eqnarray}}
\newcommand{\sph}{spin-$\frac{1}{2}$ }
\newcommand{\ad}{\hat{a}^{\dagger}}
\newcommand{\add}{\hat{a}}
\newcommand{\spp}{\hat{\sigma}_+}
\newcommand{\smm}{\hat{\sigma}_-}
\newcommand{\fin}[1]{|\phi^{\pm}_{#1}\rangle}
\newcommand{\finp}[1]{|\phi^{+}_{#1}\rangle}
\newcommand{\finm}[1]{|\phi^{-}_{#1}\rangle}
\newcommand{\lfin}[1]{\langle \phi^{\pm}_{#1}|}
\newcommand{\lfinp}[1]{\langle \phi^{+}_{#1}|}
\newcommand{\lfinm}[1]{\langle \phi^{-}_{#1}|}
\newcommand{\lfinn}[1]{\langle\phi^{\pm}_{#1}|}
\newcommand{\z}{\cal{Z}}
\newcommand{\RI}{\hat{{\cal{R}}}_{0}}
\newcommand{\Rt}{\hat{{\cal{R}}}_{\tau}}
\newcommand{\cb}{\bar{c}}
\newcommand{\nb}{\bar{n}}
\newcommand{\dnz}{ \delta n(\vec{r}_0,t)}
\newcommand{\dcz}{ \delta c(\vec{x}_0,t)}
\newcommand{\dn}{ \delta n(\vec{r},t)}
\newcommand{\dc}{ \delta c(\vec{x},t)}
\newcommand{\dch}{ \delta \hat{c}(\vec{q},q_{\perp},\omega)}
\newcommand{\dnh}{ \delta \hat{n}(\vec{q},\omega)}
\newcommand{\dnhq}{ \delta \hat{n}(\vec{q},\omega)}
\newcommand{\dnhz}{ \delta \hat{n}(\vec{r}_0,\omega)}
\newcommand{\dchz}{ \delta \hat{c}(\vec{r}_0,z_0,\omega)}
\newcommand{\nv}{  n(\vec{r},t)}
\newcommand{\cv}{ c(\vec{r},t)}
\newcommand{\nn}{\nonumber}
\newcommand{\rnb}{(\rho_0-\nb)}
\newcommand{\rnbo}{(1-\nb)}

\maketitle
\vspace{-20cm}
\tableofcontents
\setcounter{tocdepth}{20}

\newpage

\section{Berg-Purcell limit}
Consider a receptor which binds and unbinds ligand molecules with kinetics for the average occupancy $\Gamma(t)$  
given by 
\begin{equation}
\frac{d\Gamma}{dt}=k_+ c_0(1-\Gamma)\ -\ k_-\Gamma,\label{dpdt}
\end{equation}
where $k_+ c_0$ is the rate of binding at ligand concentration $ c$ and $k_-$ is the rate of unbinding. 
At steady state, \textcolor{black}{the probability of being occupied} is given by $ p= c_0/( c_0+K_D)$ with the ligand dissociation 
constant $K_D=k_-/k_+$.

What is the uncertainty $\langle\delta c^2\rangle$ in measuring ligand concentration $ c_0$? If we have
the uncertainty in occupancy, $\langle\delta \Gamma^2\rangle$, we can use error propagation and write
for the relative uncertainty in ligand concentration
\begin{equation}
\frac{\langle\delta c^2\rangle}{ c_0^2}=
\left( c\frac{\partial p}{\partial c}\right)^{-2}\langle\delta \Gamma^2\rangle\label{dc2}
\end{equation} 
with the term in bracket evaluated at $ c_0$. 

To obtain $\langle\delta \Gamma^2\rangle$ we could be tempted to use the variance of a Bernoulli random variable, 
given by $ p(1- p)$. We would thus obtain $\langle\delta c^2\rangle/{ c_0}^2=[ p(1- p)]^{-1}\geq4$ and 
hence at least $400\%$ fractional error. This instantaneous error based on a single measurement in time can formally
be derived as follows, which comes in handy later. Equation \ref{dpdt} can be linearised around the steady-state value 
by introducing $\Gamma(t)= p+\delta \Gamma(t)$ with $\delta \Gamma(t)$ 
the fluctuations and keeping only terms linear in $\delta \Gamma(t)$. This produces
\begin{equation}
\frac{d(\delta \Gamma)}{dt}=-(k_+ c_0+k_-)\delta \Gamma+\eta_{\Gamma}
\end{equation}
with $\eta_\Gamma(t)$ the fluctuating source, given by white noise with zero average. 
(That such a linearization is valid for underlying binary dynamics was shown in \cite{clausznitzer11}.) 
Subsequent Fourier transformation from the time to the frequency domain leads to
\begin{equation}
-i\omega\delta\hat{\Gamma}=-(k_+ c_0+k_-)\delta\hat{\Gamma}+\hat\eta_{\Gamma},
\end{equation}
where we applied the Fourier transforms $\Gamma(t)=\int\frac{d\omega}{2\pi}e^{-i\omega t}\delta \hat{\Gamma}(\omega)$
and  $\eta_\Gamma(t)=\int\frac{d\omega}{2\pi}e^{-i\omega t}\delta \hat\eta_{\Gamma}(\omega)$.
The power spectrum is then obtained as
\begin{equation}
\langle\delta\hat{\Gamma}(\omega)\delta\hat{\Gamma}^*(\omega)\rangle=\langle|\delta\hat{\Gamma}^2(\omega)|^2\rangle=
\frac{Q_{\Gamma}}{\lambda_{\Gamma}^2+\omega^2}\label{dp2}
\end{equation}
with noise strength $Q_{\Gamma}=\langle|\hat\eta_{\Gamma}(\omega)|^2\rangle=2k_+ c_0(1- p)$ determined from 
Poisson statistics and frequency cut-off $\lambda_{\Gamma}=k_+ c_0+k_-$. The variance is obtained by integrating 
the power spectrum over all frequencies
\begin{equation}
\langle\delta \Gamma^2\rangle=\int\frac{d\omega}{2\pi}\langle|\delta\hat{\Gamma}(\omega)|^2\rangle=
\frac{Q_{\Gamma}}{2\lambda_{\Gamma}}= p(1- p). 
\end{equation}

In contrast, Berg and Purcell (BP) considered that the receptor has some time $T$ available in order to produce 
a measurement. We expect the longer the averaging time the more accurate the measurement. Imagine a binary 
time series of occupancy $\Gamma(t)$ recorded for time $T$. The BP limit can be derived by estimating 
the average receptor occupancy $ p$ from the time-averaged value $\Gamma_T=1/T\int dt\,\Gamma(t)$ with the 
variance given by $\langle \delta \Gamma_T^2\rangle=\langle \Gamma_T^2\rangle-\langle \Gamma_T\rangle^2$.
The variance can be determined from the autocorrelation function of the occupancy, or equivalently 
the power spectrum. Specifically, the uncertainty of the occupancy $\delta \Gamma_T^2$ can be calculated by using 
the low-frequency limit of Eq. \ref{dp2}
\begin{equation}
\langle\delta \Gamma_T^2\rangle=\frac{\langle|\delta\hat{\Gamma}(\omega\approx 0)|^2\rangle}{T}=
\frac{2 p^2(1- p)}{k_+c_0 T}.
\end{equation}
When plugged into Eq. \ref{dc2}, this reproduces the BP limit
\begin{equation}
\frac{\langle\delta c^2\rangle}{ c_0^2}=\frac{2\tau_b}{T p}=\frac{2}{\bar N}\label{dc2_BP}
\end{equation}
with the average number of binding/unbinding events given by $\bar N=T/(\tau_b+\tau_u)$ with
$\tau_b=k_-^{-1}$ and $\tau_u=(k_+ c_0)^{-1}$ the average bound and unbound time intervals. 

\section{Receptor with downstream signalling}
In addition to the receptor, let us consider a signalling molecule that is produced by the ligand-bound 
receptor, characterised by occupancy $\Gamma(t)$. The kinetics for the copy number $n(t)$ of this signalling molecule 
is given by 
\begin{equation}
\frac{dn}{dt}=k\Gamma\ -\ \tau^{-1}n,\label{dndt}
\end{equation}
where $k$ times $\Gamma$ is the rate of production and $\tau$ is the lifetime of the signalling molecule. 
At steady state, the copy number is given by $\bar n=k\tau p$. Using error propagation once more,
we can write
\begin{equation}
\frac{\langle\delta c^2\rangle}{ c_0^2}=
\left(c\frac{\partial\bar n}{\partial c}\right)^{-2}\langle\delta n^2\rangle\label{dc22}
\end{equation} 
with the term in parentheses evaluated at $ c_0$. The error based on time averaging by $T$ can be derived as before. 
Equation \ref{dndt} can be linearised by introducing $n(t)=\bar n+\delta n(t)$, producing
\begin{equation}
\frac{d(\delta n)}{dt}=k\delta \Gamma-\tau^{-1}\delta n+\eta_n
\end{equation}
with $\eta_n(t)$ the fluctuating source, given again by white noise with zero average. Subsequent Fourier transforming 
from the time to the frequency domain leads to
\begin{equation}
(\tau^{-1}-i\omega)\delta\hat n=k\delta\hat{\Gamma}+\hat\eta_n,
\end{equation}
where we applied the additional Fourier transforms $n(t)=\int\frac{d\omega}{2\pi}e^{-i\omega t}\delta \hat{n}(\omega)$
and  $\eta_n(t)=\int\frac{d\omega}{2\pi}e^{-i\omega t}\delta \hat\eta_n(\omega)$.
The power spectrum is then obtained as
\begin{equation}
\langle|\delta\hat n^2(\omega)|^2\rangle=\frac{Q_n}{\tau^{-2}+\omega^2}+
\frac{Q_{\Gamma}k^2}{(\tau^{-2}+\omega^2)(\lambda_{\Gamma}^2+\omega^2)}\label{dn2}
\end{equation}
with noise strength $Q_n=\langle|\hat\eta_n(\omega)|^2\rangle=2k p$ determined from Poisson statistics. 
The variance is obtained by calculating the time-averaged low-frequency limit
\begin{equation}
\langle\delta n_T^2\rangle=\frac{\langle|\delta\hat n(\omega\approx 0)|^2\rangle}{T}=2\bar n
\left[1+\frac{\bar n(1- p)}{k_+ c_0\tau}\right]\frac{\tau}{T}.
\end{equation}
When plugged into Eq. \ref{dc22}, this produces the following limit
\begin{equation}
\frac{\langle\delta c^2\rangle}{ c_0^2}=\left[\frac{2}{\bar N_\tau}+
\frac{2}{\bar n(1- p)^2}\right]\frac{\tau}{T}=\frac{2}{\bar N} +\frac{2}{\bar n (1-p)^2} \frac{\tau}{T} \label{dc23}
\end{equation}
with $\bar N_\tau=\tau/(\tau_b+\tau_u)$ the average number of ligand binding/unbinding events in time interval $\tau$ and $\bar N$ the average number in time $T$. 
The first term in Eq. \ref{dc23} is the BP limit (cf. Eq. \ref{dc2_BP}). 
The second term in Eq. \ref{dc23} is due to Poisson-like number fluctuations in the signalling molecule and 
could be reduced for large numbers of signalling molecules.  Hence the best one can do with an equilibrium receptor is the BP limit.
%

\section{Maximum-likelihood  estimation without ligand rebinding}
Let $N$ be a fixed number of \textcolor{black}{subsequent} bound and unbound time intervals (not the average number $\bar N$ here), 
the probability (likelihood) for such a sequence of intervals 
$\vec{\tau} = (\tau_1, \cdots \tau_{N})$ is:
\BEQ
\label{probseries}
P(\vec{\tau},c) \propto e^{-k_- T_b}e^{-k_+c T_u}k\textcolor{black}{_-^N}(k_+c)^{N},
\EEQ	
\textcolor{black}{where $T_b= \sum_{i=1}^N \tau^i_b$ and $T_u=\sum_{i=1}^N \tau^i_u$ are the total bound and unbound times (with $ T_b \simeq  N  \langle \tau_u\rangle  $ and $T_u \simeq N \langle \tau_u \rangle $  for $N$ large).}
Maximising with respect to $c$ leads to: 
 \BEQ
 \label{maxlike}
 \frac{dP}{dc}=-k_+T_uP+\frac{N}{c}P=0  \;\;\to \;\; \textcolor{black}{c_{ML}}= \frac{N}{k_+T_u},
 \EEQ
\textcolor{black}{ i.e. $c_{ML}$  is the concentration value} that maximises the likelihood. The uncertainty in concentration 
measurement can be obtained from the Cram\'er-Rao bound which connects the uncertainty in concentration 
to the Fisher information. 

Assume a set of measurements $\vec{\tau}$ distributed according to $P (\vec{\tau} , c) $ from which  an unbiased estimation of the concentration $c_0$ is performed. 
In general it can be shown that given a set of \textcolor{black}{measurements} the variance for the expected value of $c$ is bound from below 
by the Fisher information, i.e.
\BEQ
\langle \delta c^2 \rangle=\langle(\textcolor{black}{\hat{c}- c_0})^2\rangle
\geq \frac{1}{I( c_0)},
\EEQ
\textcolor{black}{ with $\hat{c}$ the estimated value of the true concentration  $c_0$ and} the Fisher information $I(c)$ defined as:
\BEQ
\label{fisher0}
I(c)=-\int d\vec{\tau} \frac{\partial^2 \log P (\vec{\tau} , c)}{\partial c^2}P (\vec{\tau} , c)
\EEQ
Using Eq. (\ref{probseries}) 
 it follows that
\BEQ
 -\frac{\partial^2  \log P (\vec{\tau} , c)}{\partial c^2}=\frac{N}{c^2}, 
\EEQ
which, when inserted in Eq. (\ref{fisher0}), leads to $I(c)=N/c^2$.
In the large-$N$ limit the Cram\'er-Rao bound becomes an equality and translates into the following 
expression for the uncertainty in concentration sensing at $c_0$:
\BEQ
\label{MLerror}
\frac{\langle \delta c^2_{ML} \rangle}{c_0^2}=\frac{1}{c_0^2 I(c_0)}=\frac{1}{N}.
\EEQ
This result is two-fold lower than the BP limit. \textcolor{black}{The difference is that the maximum-likelihood (ML) estimate considers only}
the  unbound time intervals, 
as only these contain information about the ligand concentration. 

\textcolor{black}{A few comments are in order: The exact expectation value for the estimator can easily be 
derived  from Eq. (\ref{maxlike}) leading to 
\BEQ
\label{avgcML}
\langle c_{ML}\rangle=\frac{N}{k_+}\left \langle\frac{1}{T_u}\right \rangle.
\EEQ
The probability density for the variable $T_u=\sum_{i=1}^N \tau^i_u$, i.e.  of having a sequence of $N$  unbound 
time intervals, is the $N$-times convolution of the single probability density $k_+ c_0 e^{-k_+ c_0 \tau_u}$, namely
\BEQ
\psi(T_u)=(k_+c_0)^N e^{-k+c_0 T_u} \frac{T_u^{N-1}}{(N-1)!}
\EEQ
from which it follows that
\BEQ
\label{invpsi}
\left \langle\frac{1}{T_u^m}\right\rangle=\int_0^{\infty}\frac{1}{T_u^m}\psi(T_u)dT_u=\frac{(N-1-m)!}{(N-1)!} (k_+ c_0)^{m}.
\EEQ
From Eq. (\ref{invpsi}) one can easily derive  $\langle1/T_u\rangle=\frac{k_+c_0}{N-1}$, which, by means of 
Eq. (\ref{avgcML}), leads to  
\BEQ
\langle c_{ML}\rangle=c_0\frac{N}{N-1}.
\EEQ
It follows that the estimator is unbiased (i.e. $\langle c_{ML} \rangle =c_0$) only in the asymptotic limit 
of large $N$. The results obtained here and in the main text are  consistent with this limit.}
\textcolor{black}{From Eq. (\ref{invpsi})  evaluated for $m=2$  one  can derive as well the variance for the ML estimator 
\BEA
\langle \delta c_{ML}^2\rangle&=&\langle c_{ML}^2-\langle c_{ML}\rangle^2\rangle=
\frac{N^2}{k_+^2}\left\langle\frac{1}{T_u^2}\right\rangle-\langle c_{ML}\rangle^2=\frac{c_0^2}{N-2}\left(\frac{N}{N-1}\right)^2\nonumber\\
&&\simeq_{N\gg 1} \frac{c_0^2}{N}+O(1/N^2).\label{sharp}
\EEA
It follows that  the exact value for the bound on the variance of the ML estimator  is $\frac{c_0^2}{N-2}\left(\frac{N}{N-1}\right)^2$,  
which coincides with $c_0^2/N$  in the $N \gg 1$ limit  apart from terms of order $O(N^{-2})$. The limit of large $N$
is consistent  with the assumption that $T_u \gg k_{+} c_{0}, k_{-}$ implied in the integration carried out in  Eq. (\ref{invpsi}). 
Note that  one can also define the unbiased estimator $c'_{ML}=\frac{N-1}{N}c_{ML}$ for which (using Eq. (\ref{sharp}))  
one obtains a sharper bound on the variance, given by $\langle \left(\delta c'_{ML}\right)^2\rangle=c_0^2/(N-2)$.} 
\textcolor{black}{In summary, by not using the peak value of the likelihood but the mean value, we obtain an unbiased 
estimator with a slightly lower uncertainty, i.e. $1/(N-2)$ instead of $1/(N-2) [N/(N-1)]^2$.}

\section{Bayesian Cram\'er-Rao bound including a prior}
Next we consider cells which preserve a memory of previous environment\textcolor{black}{al} conditions. Specifically, the 
Bayesian Cram\'er-Rao bound \textcolor{black}{\cite{vantree1,vantree2,vantree3}}  allows us to estimate a lower bound to the variance of the expected 
value of an estimator when a prior distribution for such an estimator is known. Let us call $\hat{c}$ 
the unbiased estimator of the true concentration $ c_0$. Such a parameter is estimated based on a set 
of measurements $\vec{\tau}$ distributed according to $P(\vec{\tau},c)$. If $\lambda(c)$ is \textcolor{black}{the} known prior 
distribution of the parameter $c_0$, then it can be shown that
\BEQ
\langle \delta c^2 \rangle=\langle(\hat c- c_0)^2\rangle\label{BCR}
\geq \frac{1}{I(\lambda)+I(c_0)},
\EEQ
where \textcolor{black}{averaging on the left-hand side is conducted using the prior distribution, leading to the
reduction in uncertainty on the right-hand side.} Specifically, 
\BEQ
I(\lambda)=\int dc\lambda(c)\left[\frac{\partial \log \lambda(c)}{\partial c}\right]^2
\EEQ
is the contribution to \textcolor{black}{the} Fisher information \textcolor{black}{from} the prior distribution and
\BEQ
\label{fisher}
I(c)=\int dc\lambda(c)\int d\vec{\tau}P(\vec{\tau},c) \left[\frac{\partial \log P(\vec{\tau},c)}{\partial c}\right]^2=-\int dc\lambda(c)\int d\vec{\tau} P(\vec{\tau},c)\frac{\partial^2 \log  P(\vec{\tau},c)}{\partial^2 c}
\EEQ
is the Fisher information about the parameter $c_0$ given the data $\vec{\tau}$.
The second equality  in  Eq. (\ref{fisher}) follows  from the relation $\frac{\partial^2 \log P(\vec{\tau},c)}{\partial^2 c}=\frac{P^{\prime \prime}}{P} -\frac{{P'}^2}{P^2}=\frac{P^{\prime \prime}}{P}-\left[\frac{\partial \log P(\vec{\tau},c)}{\partial c}\right]^2$
and that the term $\frac{P^{\prime \prime}}{P}$ gives zero contribution as can be checked by differentiating with respect to $c$ the normalisation condition
\BEQ
\int d\vec{\tau}P(\vec{\tau},c)=1.
\EEQ

\subsection{Log-normally distributed prior }
We first consider the case where the prior distribution has a log-normal form, i.e.:
\BEQ
\lambda(c)=\frac{1}{\sigma \sqrt{2 \pi} \textcolor{black}{c} }\exp\left[-\frac{(\log(c) -\mu)^2}{2\sigma^2}\right]
\EEQ
with mean and variance in log-space given by $\langle \log(c)\rangle=\mu$ and 
$\langle [\log(c)-\mu]^2\rangle=\sigma$, respectively. In linear space these
are given by respective expressions
\begin{subequations}
\BEA
\langle c\rangle&=&\exp\left(\mu+\sigma^2/2\right)\\
\langle c^2 -\langle c\rangle^2 \rangle&=& \exp[2(\mu+\sigma^2)]-\exp\left(2 \mu+\sigma^2\right)=\exp\left(2 \mu+\sigma^2\right) \left[\exp\left(\sigma^2\right)-1\right].
\EEA
\end{subequations}
Consequently, $I(\lambda)$ is given by:
\BEA
\nn I(\lambda)&=&\int dc\lambda(c)\left[\frac{\partial \log \lambda(c)}{\partial c}\right]^2=\int \frac{dc}{c_0^2}\left[\frac{1}{\sigma^2}(\log(c)-\mu)+1\right]^2 \exp{\left[-\frac{(\log(c) -\mu)^2}{2\sigma^2}\right]}\\
 &=&\left(\frac{1}{\sigma^2}+1\right)\exp{\left[-2(\mu-\sigma^2)\right]}
\EEA
Furthermore, $I(c)$  follows from Eq. (\ref{fisher0}) 
where  $P(\vec{\tau},c)$, the probability (likelihood) of observing a sequence $\vec{\tau}$ of $N$ 
bound and unbound time intervals, is given by Eq. (\ref{probseries}). Equations 
(\ref{probseries}) and (\ref{maxlike}) lead to the following expression for $I(c)$:
\BEQ
\label{baysFisher}
I(c)=\int dc\int  d\vec{\tau} P(\vec{\tau},c)\frac{N}{c^2} \lambda(c).
\EEQ
\textcolor{black}{Performing}
 the integration in $\vec{\tau}$ due to the normalisation condition,
we obtain:
\BEQ
I(c)=\int dc\frac{N}{c^2} \lambda(c)=N \exp[-2(\mu-\sigma^2)].
\EEQ
In conclusion, the uncertainty in ligand concentration with the Bayesian Cram\'er-Rao bound and a 
log-normal prior is given by
\BEQ
\label{lognvar}
\langle\delta c^2\rangle\geq \frac{\exp[2(\mu-\sigma^2)]}{N+1/\sigma^2 +1}.
\EEQ

\textcolor{black}{In the following we deviate slightly from the derivation found in \cite{aquino14}.
In this article the prior was assumed to be centred around the true ligand concentration. Here,
we assume more conservatively that the prior distribution is centred around the 
(erroneous) ML value $c_{ML}$ of the concentration obtained from the previous measurement. 
The variance of the distribution is again given by the standard Cram\'er-Rao bound from the ML 
estimation, $\langle\delta c_{ML}^2\rangle=c_0^2/N$ \cite{end09} with $c_0$ the true value for 
the concentration.} From these assumptions it follows that
\begin{subequations}
\label{meansdvrep}
\BEA
\label{meansdvrepa}
\langle c\rangle&=&\exp{\left[\mu+\frac{\sigma^2}{2}\right]}=c_{ML}\\
\label{meansdvrepb} \langle c^2- \langle c \rangle^2 \rangle &=&\left(\exp(\sigma^2)-1\right) \langle c\rangle^2 =\frac{c_0^2}{N},
\EEA
\end{subequations}
where last equality follows from Eq.  (\ref{MLerror}).  We can now  express $\sigma^2$ in terms of known quantities, noticing that, for  large number of events $N$, from Eq. (\ref{meansdvrepb})  it follows:
\BEQ
\exp[\sigma^2]-1 \textcolor{black}{\simeq }\frac{1}{N} \;\; \;\to \; \; \; \sigma^2\textcolor{black}{\simeq }\log\left(1+\frac{1}{N}\right)\simeq \frac{1}{N}.
\EEQ
Inserting  this expression for $\sigma^2$ into  Eq. (\ref{lognvar}) leads to \textcolor{black}{$\langle\delta c^2\rangle\geq \frac{c_{ML}^2 e^{-3/N}}{N+1/N +1}$
and therefore to}
\BEQ
\label{lognRes}
\frac{\langle \delta c^2\rangle}{ c_0^2}\geq \frac{1}{2 N},
\EEQ
where $c_0$ in the denominator is the true value of the concentration, 
which differs from $c_{ML}$ obtained from 
previous measurement by  at most a correction proportional to $c_0/\sqrt{N}$ (due to Eq. (\ref{MLerror})), 
 so that Eq. (\ref{lognRes}) is correct to 
leading order for $N$ large, \textcolor{black}{ a result identical to the one in \cite{aquino14}.} 

Eq. (\ref{lognRes})  means that having a prior distribution for $N$ intervals is the same as measuring for $2N$ 
intervals without a prior distribution. Hence, information is neither lost nor gained. 
This also means that by using memory (in \textcolor{black}{the} form of a prior) a cell can effectively perform longer and hence 
more accurate measurements \textcolor{black}{ 
without being limited  by  the actual measurement (averaging) time.} 

\subsection{Gamma-distributed prior}
Alternatively, assume the prior is given by the Gamma distribution
\BEQ
\lambda(c)=\frac{c^{\alpha-1} \gamma^{\alpha}e^{-\gamma c}}{\Gamma[\alpha]},
\EEQ
where the parameters $\alpha$ and $\gamma$ are related to the 
first and second moment of the distribution:
\begin{subequations}
\BEA
\langle c\rangle&=&\frac{\alpha}{\gamma}\label{cma}\\
\langle c^2-  \langle c\rangle)^2\rangle&=&\frac{\alpha}{\gamma^2} \label{cmb}.
\EEA
\end{subequations}
For such a  prior distribution, $I(\lambda)$  reads:
\BEA
\label{ref25}
I(\lambda)&=&\int \left[\frac{\partial \log \lambda(c)}{\partial c}\right]^2 \lambda(c) dc=\\ \nonumber
&&\int \frac{\lambda^{\prime}(c)^2}{\lambda(c)}dc=\int_0^\infty dc\frac{e^{-\gamma c} (\gamma c)^{\alpha+1}
(\alpha-1-\gamma c)^2}{\gamma c^4 \Gamma[\alpha]}=\frac{\gamma^2}{\alpha-2}.
\EEA
The Fisher information $I(c)$ is determined from \textcolor{black}{Eq. (\ref{baysFisher})} 
 using the Gamma distribution instead of the log-normal distribution
\BEA
I(c)&=&\int dc\int  d\vec{\tau} P(\vec{\tau},c)\frac{N}{c^2} \lambda(c)=\\ \nonumber
&&\int \frac{N}{c^2} \lambda(c) dc=N \gamma^2  \int_0^{\infty}dc\frac{c^{\alpha-3}
e^{-\gamma c}\gamma^{\alpha -2}}{\Gamma[\alpha]}=\frac{N \gamma^2}{(\alpha-1)(\alpha-2)}.
\EEA
Consequently, the Bayesian Cram\'er-Rao bound is given by:
\BEQ
\langle\delta c^2\rangle\geq \frac{1}{\frac{\gamma^2}{\alpha-2}+\frac{N \gamma^2}{(\alpha-1)(\alpha-2)}}.
\EEQ
With the same assumptions as done in the previous section, 
mean value and standard deviation of the prior distribution are set
 to $ c_{ML}$ and $  c_{ML}^2/N$, respectively (see Eqs. (\ref{meansdvrep})), 
\textcolor{black}{with $ c_{ML}$ the ML value for the concentration obtained in previous measurement. Eqs. (\ref{cma}) and (\ref{cmb})  then imply}:
\begin{subequations}
\BEA
\gamma&=&\frac{N}{ c_{ML}}\\
\alpha&=&N.
\EEA
\end{subequations}
In the limit of large $N$, this leads to
relative uncertainty
\BEQ
\frac{\langle\delta c^2\rangle}{c_0^2}\geq \frac{1}{\frac{N^2}{(N-2)}+
\frac{N^3}{(N-1)(N-2)}}\simeq \frac{1}{2 N},
\EEQ
\textcolor{black}{
with again $c_0$ the true value of the concentration
($ c_{ML}\simeq c_0 \pm c_0/\sqrt{N}$).}
Not surprisingly, this is the same result as obtained in Eq. (\ref{lognRes}), since
both  the log-normal and the Gamma distribution can be derived from Gaussian distributed variables. The Gamma distribution 
is the distribution of the sum of squared normal variables (a.k.a.$\chi^2$ distribution), while the log-normal, as the  
name suggests, is the distribution of the logarithm of a normally distributed variable.

\section{Maximum-likelihood estimation with ligand rebinding}

Endres and Wingreen \cite {end09} applied maximum likelihood (ML) to \textcolor{black}{the problem of } estimating the ligand 
concentration from a time series of ligand-receptor occupancy, but focused on the uncertainty of this measurement 
without ligand \textcolor{black}{rebinding}, i.e. effectively for very fast diffusion. \textcolor{black}{For slower 
diffusion one should consider}  possible rebinding of a previously bound ligand molecule, which makes the 
instantaneous rate of binding a functional of the previous binding and unbinding events. The binding 
rate can thus be written as $k_+c_0(t,\{t_{+},t_{-}\})$. The rate of unbinding remains $k_-$, so the ML 
estimate of concentration still comes entirely from the durations of the unbound intervals. 

We quickly review ML estimation of the ligand concentration with ligand rebinding \cite{end09}.  The probability for a time series to occur given a ligand concentration $c_0$ is
\begin{equation}
P(\{t_{+},t_{-}\};c)=\prod_i p_b(t_{+,i},t_{-,i})p_-(t_{-,i})p_u(t_{-,i},t_{+,i+1})p_+(t_{+,i+1}),\label{eq:P1}
\end{equation}
where the probability for a ligand molecule to remain bound from $t_{+,i}$ to $t_{-,i}$ is 
\begin{equation}
p_b(t_{+,i},t_{-,i})=p_b(t_{-,i}-t_{+,i})=e^{-k_-(t_{-,i}-t_{+,i})}.
\end{equation}
 \textcolor{black}{The probability for a receptor to remain unbound from $t_{-,i}$ to $t_{+,i+1}$
 includes the effect on the binding of the changing concentration of ligand $p_+\propto k_+(c_0+\Delta c_i)$  where $\Delta c_i$ is the perturbation to the ligand concentration from previous binding and unbinding events. Consequently}
\begin{equation}
p_u(t_{-,i},t_{+,i+1})=e^{-k_+c_0(t_{+,i+1}-t_{-,i})-k_+\int_i \Delta c(t')dt'},\label{eq:pu}
\end{equation}
where we have expressed the ligand concentration as
\begin{equation}
c(t,\{t_+,t_-\})=c_0+\Delta c(t,\{t_{+},t_{-}\})=c_0+\Delta c(\{t-t_{-,i};t-t_{+,i}\}),
\end{equation}
and used the notation $\int_idt'=\int_{t_{-,i}}^{t_{+,i+1}}dt'$,
$\Delta c(t')=\Delta c(t',\{t_+,t_-\})$, and $\Delta c_i=\Delta c(t_{+,i})$.

The terms can be gathered as before, leading to
\begin{equation}
P(\{t_{+},t_{-}\};c)\propto e^{-k_-T_b}\cdot e^{-k_+c_0T_u}\cdot k_-^{N}\cdot k_+^{N}
\cdot\prod_i(c_0+\Delta c_i)e^{-k_+\int_i\Delta c(t') dt'}.\label{eq:P2}
\end{equation}
Importantly, all the $\Delta c$'s depend only on the times of events, not the value of $c_0$, so 
$d(\Delta c)/dc=0$, yielding
\begin{equation}
\frac{dP}{dc}\propto-k_+T_uP+\sum_i\frac{1}{c_0+\Delta c_{i}}P.
\end{equation}
Setting the above derivative to zero yields an implicit equation for the ML estimate of $c_0$,
\begin{equation}
\sum_i\frac{1}{c_0+\Delta c_i}=k_+T_u,\label{eq:c_ML}
\end{equation}
where the sum is over all binding events. Importantly, each $\Delta c_i$ depends deterministically 
on all previous binding and unbinding events. For \textcolor{black}{the} special case  \textcolor{black}{of fast diffusion} $D=\infty$ and hence $\Delta c_i=0$, 
we obtain \cite{end09} $k_+c_0=N/T_u=1/\langle\tau_u\rangle$ where $T_u$ is the total unbound
time of the receptor during time $T$, $N$ is the total number of binding/unbinding events, and 
$\langle\tau_u\rangle$ is the average unbound time interval.

How accurate is the concentration estimate? Using the Cram\'er-Rao bound once more, we obtain for the normalised variance
\begin{equation}
\frac{\langle\delta c^2\rangle}{c_0^2}=-\frac{1}{c_0^2\left\langle\frac{d^2\ln(P)}{dc^2}\right\rangle_{c_0}}
=\frac{1}{\langle\sum_i(1+\Delta c_i/c_0)^{-2}\rangle_{c_0}},\label{eq:ML_D}
\end{equation}
where we used $P$ from Eq. \ref{eq:P2}.
Hence, the normalised variance of the ML estimate of the true concentration $c_0$ is
the inverse of the number of unbound intervals with additional corrections  \textcolor{black}{in the regime of slow diffusion}, due to  perturbations in
ligand concentration  \textcolor{black}{from previous binding and unbinding events.}

Equation \ref{eq:ML_D} depends on the average over all trajectories with $N$ binding and unbinding events.
Furthermore, each perturbation in ligand concentration, $\Delta c_i$, depends on the whole history of 
binding and unbinding events,  making this equation unsolvable. However, we can estimate the effect
of diffusion in the limit of slow binding and unbinding, or fast diffusion. Hence, for small $\Delta c_i/c$, 
we can expand to linear order
\begin{equation}
\frac{\langle\delta c^2\rangle}{ c_0^2}\approx
\frac{1}{N}\left(1+2\frac{\langle\Delta c\rangle}{c_0}\right)\label{dc2ML}
\end{equation}
to simplify the equation for the uncertainty. Equation \ref{dc2ML} now contains only a typical perturbation 
in ligand concentration $\langle\Delta c\rangle$. To estimate this we use the solution of the diffusion
equation for a single ligand molecule
\begin{equation}
\Delta \textcolor{black}{_{\pm}} c(\vec r,t)=\frac{\pm1}{(4\pi Dt)^{d/2}}e^{-\frac{r^2}{4Dt}},\label{Dcd}
\end{equation}
with $+1$ corresponding to an unbound ligand molecule (source) and $-1$ corresponding to a bound
ligand molecule (sink) at $t=0$. We assume the receptor sits at $\vec r=0$, at which we wish to
evaluate perturbation. Here, we provide results for dimensions $d=2$ and $3$.\\

\noindent {\bf 2-dimensional diffusion:} Here we consider $d=2$ in Eq. \ref{Dcd}. To further  \textcolor{black}{simplify} the calculation of the 
whole history of binding and unbinding events, we assume all binding events are independent, i.e. 
only depend on  \textcolor{black}{the} average rate $k_+ c_0$ and not also on the perturbations. We thus obtain the infinite series
\begin{equation}
\label{sumtricky}
\langle\Delta c (t) \rangle=\frac{1}{4\pi D}\left(\left\langle\frac{1}{\tau_1}\right\rangle
-\left\langle\frac{1}{\tau_1+\tau_2}\right\rangle +... (-1)^{K+1}\left\langle\frac{1}{\tau_1+\tau_2+...+\tau_K}\right\rangle  +...\right),
\end{equation}
with the most recent  (unbinding)  event occurring  at time $t-\tau_1$ and increasing the overall concentration (source) and the second most recent (binding) event occurring at time 
$t-(\tau_1+\tau_2)$ and decreasing the overall concentration (sink),  and so on. The averages are performed over the probability of a sequence of $K$ events and then summed  in the limit $K\to \infty $ to account for an infinitely long history.


  For the special case $ p=1/2$, so that $\langle \tau_u\rangle =\langle \tau_b\rangle=\tau$ and  $\lambda=k_+ c_0=k_-$,  each random number $\tau$ is generated with the same distribution $\psi(\tau)=\lambda e^{-\lambda \tau}$.
    \textcolor{black}{  In order to evaluate the generic term in the series of Eq. (\ref{sumtricky}) one has first
 to evaluate the probability density that a given value $T_K$ is obtained for the sum $\sum_{i=1}^K \tau_i=T_K$ after $K$  draws of the random variable $\tau$. This probability is given by the $K$-times convolution of the   distribution  $\psi(\tau)$, which is:
 \BEQ
 \label{conv}
 \psi_K(T_K)= \lambda^{K} e^{-\lambda T_K} \frac{T_K^{K-1}}{(K-1)!}. 
 \EEQ
Then, by using this distribution one can evaluate $\langle \frac{1}{T_K}\rangle$.
 This allows us to obtain an expression for a generic term  with $K\geq 2$ in the sum in Eq. (\ref{sumtricky}),  leading to
 \BEQ
\left  \langle \frac{1}{T_K} \right \rangle=\int_0^{\infty} \frac{1}{T_K} \psi_K(T_K) dT_K=\frac{\lambda}{K-1}.
 \EEQ
Summing all these contributions for $K\geq2$ leads to
\BEQ
\label{Km2}
\sum_{K=2}^{\infty}(-1)^{K+1}\frac{\lambda}{K-1}=-\lambda \log 2.
\EEQ
The contribution for $K=1$ has to be calculated separately. In fact, one has to} calculate $\langle 1/\tau\rangle=
\lambda\int_0^\infty dt\,e^{-\lambda t}/t$,
which is the Gamma function $\Gamma(n)$ for diverging parameter $n=-1$.
To make progress we realise that the maximal perturbation is the change in concentration 
 \textcolor{black}{due to a single
ligand molecule released into a $2D$ area of the order of the size of the binding site, so $\Delta c_{+,max}\simeq 1/a^2=\frac{1}{4 \pi D \tau_a}$}. This effectively introduces a minimal time $\tau_a=a^2/(4 \textcolor{black}{ \pi }D)$.
As a result, we approximate the integral by
\begin{equation}
\left\langle\frac{1}{\tau}\right\rangle=\lambda\int_{\tau_{\textcolor{black}{a}}}^\infty dt\frac{e^{-\lambda t}}{t}=-\lambda{\rm Ei}(-x)
=-\lambda\left[\gamma+\ln x+\sum_{k=1}^\infty (-1)^k\frac{x^k}{k(k!)}\right]
\end{equation}
with $x=k_+c_0\tau_a<\!\!<1$ and hence $\ln x<\!\!<0$, and $\gamma\approx 0.57721...$ the Euler-Mascheroni constant. 
For very small $x$, $\gamma$ and the sum can be neglected  and the dominant term is the logarithmic part. In this limit it is evident as well that
the contribution of  the most recent event is
 much larger than the contribution from all other events in Eq. (\ref{Km2}), so that the final result is
 \textcolor{black}{ 
\begin{equation}
\langle\Delta c\rangle\approx\frac{k_+c_0}{4\pi D}\ln\left(\frac{4\pi D}{k_+c_0a^2}\right),
\end{equation} }
showing the competition between rebinding with rate $k_+ c_0$ and diffusion to remove the unbound ligand molecule.
\textcolor{black}{As expected for 2D, this result only shows a weak dependence on diffusion and receptor size.}
\\

\noindent {\bf 3-dimensional diffusion:} Here we set $d=3$ in Eq. \ref{Dcd} and obtain
\BEQ
\label{sum3D}
\langle \Delta c\rangle=\frac{1}{(4 \pi D)^{3/2}} \left( \langle \tau_1^{-3/2} \rangle- \langle(\tau_1+\tau_2)^{-3/2}\rangle+... (-1)^{K+1}\langle(\tau_1+\tau_2+...+\tau_K)^{-3/2} \rangle ... \right).
\EEQ
The calculation therefore,  under the  simplifying assumption  $\lambda=k_+ c_0=k_-$, is analogous to the 2-dimensional case with the only difference that the average  $\langle \tau^{-3/2}\rangle=\lambda \int_0^\infty d\tau e^{-\lambda \tau}/\tau^{3/2}$ replaces  $\langle \tau^{-1}\rangle$. Also in this case  it can be shown that the contribution from the most recent event is dominant as compared to that of all other events (i.e. terms in Eq. (\ref{sum3D}) with $K\geq 2$). \textcolor{black}{ 
  Using  the distribution  in Eq. (\ref{conv})  it is possible to calculate the contributions from all the terms with $K\geq 2$ in Eq. (\ref{sum3D}). One obtains for a generic term in the sum:
\BEQ
\left \langle \frac{1}{T_K^{3/2}} \right \rangle=\int_0^{\infty} T_K^{-3/2} \psi_K(T_K) dT_K=\lambda^{3/2} \frac{\Gamma[K-3/2]}{\Gamma[K]},
\EEQ
which, after summation, leads to:
\BEQ
\label{3DKm2}
\sum_{K=2}^{\infty}(-1)^{K+1}\left\langle \frac{1}{T_K^{3/2}}\right \rangle=-2(\sqrt{2}-1)\sqrt{\pi} \lambda^{3/2}.
\EEQ
The contribution from the most recent event (i.e. first term in Eq. (\ref{sum3D})) is given by $\langle \tau^{-3/2}\rangle=\lambda \int_0^\infty d\tau e^{-\lambda \tau}/\tau^{3/2}$.
Introducing $x=k_+ c_0 \tau_a=\lambda \tau_a \ll 1$ as in the 2D case, the evaluation of $\langle \tau^{-3/2}\rangle$ leads to
\BEQ
\langle{\tau}^{-3/2}\rangle=\lambda \int_{\tau_a}^\infty d\tau e^{-\lambda \tau}/\tau^{3/2}=2\lambda^{3/2} \left[ \frac{e^{-x }}{\sqrt{x}} 
-  \sqrt{ \pi} \text{Erf} \left(\sqrt{x}\right)\right],
\EEQ
which, to leading order in $x<\!\!<1$,  leads to
\BEQ
\langle \tau^{-3/2}\rangle \simeq \frac{2 \lambda}{\sqrt{a^2/(4 \pi D)}}.
\EEQ
The ratio between the contribution from terms with $K\geq2$ of Eq. (\ref{3DKm2}) and this contribution 
amounts to $\sim \sqrt{\lambda \tau_a}=\sqrt{x} \ll 1$, which justifies keeping
only the first term in Eq. (\ref{sum3D}) in the limit $x\ll1$.}
The final result is therefore 
\BEQ
\langle \Delta c\rangle\simeq\frac{1}{(4 \pi D)^{3/2} }\langle{\tau}^{-3/2}\rangle\simeq\frac{1}{(4 \pi D)^{3/2} }
\frac{2k_+ c_0}{\sqrt{a^2/(4 \pi D)}}=\frac{k_+ c_0}{2 \pi D a},
\EEQ
\textcolor{black}{which has a stronger dependence on diffusion and receptor size compared to sensing in 2D.}
The results for diffusion in 2D and 3D are stated in the main text. Specifically, Eq. 12 in the main text is 
obtained by using $N=T_u/\langle \tau_u\rangle$ for the number of binding/unbinding intervals
with $T_u=(1-p)T$ and $\langle \tau_u\rangle=(k_+c_0)^{-1}$. \textcolor{black}{Note that while we estimate $\langle\Delta c\rangle$ 
from the whole history of binding events, in our calculation the individual binding events depend only on the average ligand concentration
$c_0$. Hence, similar to other derivations of the uncertainty of sensing by a receptor with ligand rebinding (main text 
Eqs. (10) \cite{bialek05} and (11) \cite{kaizu14}), 
we calculate the first-order correction to the uncertainty due to ligand diffusion and rebinding. Note also that only
the last unbinding event counts for deriving $\langle\Delta c\rangle$ so the exact durations of former unbound intervals 
do not matter.}

\section{Uncertainty and decision-making algorithms}
Rapid and accurate decisions are ubiquitously made in cells  \textcolor{black}{motivating modelling of } how this timely accuracy is achieved.
Here, we follow Siggia and Vergassola  to evaluate the decision time and  uncertainty associated with 
optimal  \textcolor{black}{decision-making}  algorithms  \cite{siggia13}. In particular, we want to make a connection between decision-making algorithms and
ML estimation/BP limit.

In the case of deciding between two options, e.g. two concentration values  \textcolor{black}{$c_1$ and $c_2$},
it can be shown that the Wald algorithm is, on average, optimal in time. Given a  \textcolor{black}{ fixed   probability 
 that
the  wrong decision is made, the  Wald
algorithm makes the decision in the shortest amount of time on average.  In this algorithm two   fixed thresholds 
$H_1$ and \textcolor{black}{$H_2>H_1$} are given, and  at each time step} 
the ratio  
\begin{equation}
\label{Rdef}
R=L(\text{data}|c_1)/L(\text{data}|c_2)
\end{equation}
 \textcolor{black}{between the likelihoods conditioned to either option is evaluated}. Concentration  $c_1 (c_2)$ is chosen if $R\leq H_1 (R\geq H_2)$, while 
data acquisition continues if  $H_1<R<H_2$. The algorithm can be mapped to  a diffusion process of the 
variable  $\ln R$  \textcolor{black}{between} two absorbing boundaries, corresponding to the thresholds, 
the values of which are in turn directly connected to the
decision-error  probability that  $c_{1} (c_2)$ is wrongly chosen when real value is  $c_{2} (c_1)$. 
In the diffusive approximation for $\ln R$ the average absorption time $\langle T_{\text{abs}} \rangle$, 
which coincides with the decision time, is given by \cite{siggia13}
\begin{equation}
\label{tabs}
\langle T_{\text{abs}}\rangle=\frac{x}{V}+ \frac{K}{ V \sinh (VK/D)} \left[\cosh(KV/D)-e^{-x V/D} \right]
\end{equation}
with $x$ the initial value,  $V$ the drift and  $D$ the diffusivity of $\ln R$, and symmetric absorbing  
boundaries at $x=\pm K=\pm \frac{1}{2}\log(\frac{H_2}{H_1})$ .
 
\section{Neyman-Pearson lemma}
In the Wald algorithm time is not constrained to be fixed, and this algorithm is optimal in time on average. 
If time is constrained, i.e. fixed sample or data size, the optimal test is given by the Neyman-Pearson (NP) lemma.
When choosing between two options  $c_1$ (reference hypothesis) and $c_2$  with a criterion A for rejecting  $c_1$ and a given
probability of a decision error
\begin{equation}
\label{nym1}
\alpha=P(A |c_1), 
\end{equation}
i.e. of wrongly choosing $c_2$ when the data are generated with $c_1$, 
the optimal choice is made by rejecting $c_1$ in favour of  $c_2$ if $R\leq H$ and choosing $c_1$ otherwise.
Again, $R$ is the likelihood ratio of Eq. (\ref{Rdef})  and $H$ is an $\alpha$-dependent threshold (NP lemma).  
This optimal criterion fulfils by definition  the constraint of Eq. (\ref{nym1})
\begin{equation}
\label{nym2}
\alpha=P(R\leq H |c_1)\textcolor{black}{=\int_{x:R\leq H}L(x|c_1)dx},
\end{equation}
which shows also that the threshold value $H$ is determined by
 \textcolor{black}{ $\alpha$}. This algorithm is optimal in the sense 
that any other algorithm  based on a different rejection criterion   $A$  of the reference hypothesis $c_1$  but with the same 
$\alpha$,  will have a smaller  probability $P(A |c_2)$ of correctly choosing $c_2$ (i.e. correctly rejecting  $c_1$) as compared to  the analogous NP's   probability  $P(R\leq H |c_2)$:
\BEQ
P(R\leq H |c_2)\textcolor{black}{=\int_{x:R\leq H}L(x|c_2)dx} \geq P(A |c_2)\textcolor{black}{=\int_{x: A}L(x|c_2)dx} \;\;\;\;\;\; \forall A\;\; \text{on}\;\;x.
\EEQ
 In other words for all possible data generated 
with $c_2$ the NP test will correctly choose $c_2$ more often than any other test with the same 
 \textcolor{black}{$\alpha$}.  
For this reason this algorithm may be seen as a maximal-likelihood decision-making algorithm as the likelihood of 
correctly choosing $c_2$ is maximal. So while the Wald algorithm is optimal in time on average, for a 
fixed time the NP algorithm leads to a maximum likelihood of the correct decision.

\section{Decision-making algorithms vs.  Berg-Purcell limit }

Berg and Purcell (BP) were the first to derive an estimate for the uncertainty of a measurement of ligand concentration 
in a fixed time $T$ by a  small detecting device, e.g. a cell. Here we consider their model as applied to a 
single receptor. As discussed, their estimate was later improved  by  the maximum-likelihood (ML)  estimate, 
showing that the uncertainty is actually smaller by a factor two because only unbound intervals carry 
information about the external concentration \cite{end09}. Estimates for the uncertainty refer to 
concentration measurements and cannot directly be compared to decision making between two values.

However, one can  still assume that in the BP and ML estimates, different concentrations 
$c_1$ and $c_2$ can be told apart if $\sqrt{\langle\delta c^2\rangle} < |c_2 -c_1|$ and that, assuming either 
$c_1$ or $c_2$ as the true value, a decision error occurs if a measurement returns a value outside one standard 
deviation from the true value. With these choices for the thresholds one accounts for large fluctuations 
around the true value, which may lead to wrongly deciding on the other of two possible options for the true value.
We are now in a position in which we can attempt to compare BP and ML estimates with the Wald and NP algorithms.  
Setting the decision error $\alpha$ of the Wald and NP algorithms equal to the decision error from both 
the BP and ML estimates introduced above, one obtains minimum value $|c_2-c_1|$ for fixed value $c_1$ so that a decision 
between $c_2$ and $c_1$ can be made with error $\alpha$. This value can in turn be used as a definition of 
the uncertainty in decision making and compared to the BP and ML estimates. 

Specifically, for a given value of $T$ for the Wald algorithm one can use Eq. (\ref{tabs}) and, 
for given $c_1$, derive  the corresponding  value of $c_2$ that can be distinguished in time $T$ with 
decision error $\alpha=\alpha(K)$, with $x=\pm K$ the absorbing boundaries for the symmetric case. 
Using this approach, we can plot $(c_2-c_1)^2$ of the Wald algorithm as a function of $T$ and hence 
indirectly compare to the uncertainties from BP and ML (Fig. 5). A similar procedure allows us to extract 
$(c_2-c_1)^2$ for the fixed-time Neyman-Pearson lemma (which is also shown in {Fig. 5}).

\end{document}